\documentclass[11pt]{article}

\input{tcilatex}

\begin{document}

\date{}
\title{Equivalence of Many-Photon Green Functions in DKP and KGF Statistical
Quantum Field Theories.}
\author{V.Ya.Fainberg\thanks{%
P.N. Lebedev Physical Institute, Moscow, Russia.{}}, B.M. Pimentel\thanks{%
Instituto de F\'{\i}sica Te\'{o}rica, Universidade Estadual Paulista,
S\~{a}o Paulo, SP, Brazil.} and J. S. Valverde\footnotemark[2]  \\
{\small \textit{\ }}\\
}
\maketitle

\begin{abstract}
We prove the equivalence of many-photon Green functions in statistical
quantum field Duffin-Kemmer-Petiau (DKP) and Klein-Gordon-Fock (KGF)
theories using functional path integral formalism for partition functional
in statistical quantum (finite temperature) field theory. We also calculate
the polarization operators in these theories in one-loop approximation, and
demonstrate their coincidence.
\end{abstract}

\section{Introduction}

The method of Green functions (GF) in quantum statistics has a long history
which begins with Matzubara's work in 1953 \cite{Matzubara}. The method of
generating or partition functional was first applied for calculation of
temperature renormalizable GF in \cite{Fradkin}, \cite{Proc} by Fradkin%
\footnote{%
The detailed list of publications may be found in Proceedings of P.N.Lebedev
Physical Institute, vol. \uppercase\expandafter{\romannumeral 29}, 1965, in
Fradkin's dissertation \cite{Proc}, and in Kapusta's \cite{Kapusta} and Le
Bellac's \cite{Le Bellac} books. See also references in the paper \cite
{Casana}.}. Later, the method of functional path-integral in statistics was
developed in Bernard's work \cite{Bernard}.

In the last years, the equivalence of DKP and KGF theories was proved in 
\cite{PF}, \cite{FP}, \cite{FP1} for the mass shell S-matrix elements of
scalar-charged particles interacting with the quantized EM and YM fields, as
well as for GF with external photons \textit{off the mass shell}.

An interesting physical question arises in this connection: can one prove
the equivalence of many-photon GF in statistical quantum (finite
temperature) DKP and KGF theories?

The main goal of this paper is to give positive answer to this question.

In section \textbf{2} we give the general proof of the equivalence by path
integral method in statistical quantum theory. This result can also be
understood from the physical point of view. Photon does not acquire mass and
consequently the chemical potential $\mu $ too; i.e., photon in temperature
medium conserves its nature. As an illustration, in section \textbf{3} we
calculate polarization operators in one-loop approximation of both theories,
and prove that these operators do coincide.

Section \textbf{4} contains the conclusions.

\section{Coincidence of Many-Photon GF in DKP and KGF at Finite Temperature
Theories}

To obtain the partition functional $Z(J,\bar{J},J_{\mu })$ in statistical
theory one must make transition to Euclidean space and restrict integration
on $x_{4}$: $0\leq x_{4}\leq \beta $; here $\beta =1/T$ and $J,\bar{J}%
,J_{\mu }$ are external currents. As it\ follows from general
considerations, the partition functional in DKP theory of charged spin-zero
particles interacting with the quantized EM field $A_{\mu }$ (in $\alpha $%
-gauge) has the following form\footnote{%
See the paper \cite{Bernard} for the discussion of gauge-dependence.}: 
\begin{eqnarray}
&&Z_{DKP}=Z_{0}\int_{\beta }\,DA_{\mu }\,D\psi \,D\bar{\psi}\,\exp \left\{
-\int_{0}^{\beta }dx_{4}\int_{-\infty }^{\infty }d\mathbf{x}\left[ \frac{1}{4%
}F_{\mu \nu }^{2}+\frac{1}{2\alpha }(\partial _{\mu }A_{\mu })^{2}\right.
\right.  \nonumber \\
&&\qquad {}\left. \left. +\bar{\psi}(x)(\beta _{\mu }D_{\mu }+m)\psi
(x)+J_{\mu }(x)A_{\mu }(x)+\bar{J}(x)\psi (x)+\bar{\psi}(x)J(x)\right]
\right\} ,
\end{eqnarray}
where $Z_{0}=Z(0,0,0)$\footnote{%
In the case of charge $e=0$: 
\[
Z_{0}=\prod_{\mathbf{p},\mathbf{k}}\left( 1-\exp (-\beta (E-\mu ))\right)
^{-1}\left( 1-\exp (-\beta \omega )\right) ^{-1}, 
\]
where $E=\sqrt{\mathbf{p}^{2}+m^{2}},$ $\omega =|\mathbf{k}|$.}; $D_{\mu
}=\partial _{\mu }^{\ast }-ieA_{\mu },$ $\partial _{4}^{\ast }=\partial
_{4}-\mu $; $\mu $ is the chemical potential\footnote{%
In Bose-Einstein case the statistical potential is always negative.}. In
Euclidean space $\bar{\psi}(x)=\psi ^{\ast }(x)$; all the fields satisfy
periodical conditions: 
\begin{equation}
\bar{\psi}(0,\mathbf{x})=\bar{\psi}(\beta ,\mathbf{x}),\quad {\psi }(0,%
\mathbf{x})={\psi }(\beta ,\mathbf{x}),\quad A_{\mu }(0,\mathbf{x})=A_{\mu
}(\beta ,\mathbf{x}).
\end{equation}
For example, in Eq.~(1): 
\begin{equation}
\int_{\beta }DA_{\mu }(x)=\int \prod_{0\leq x_{4}\leq \beta }\prod_{-\infty
\leq x_{i}\leq \infty }\prod dA_{\mu }(x_{4},\mathbf{x}).
\end{equation}
In Euclidean space $\bar{\psi}(x)=\psi ^{\ast }(x)$; we choose $\beta _{\mu
} $-matrices in the form: 
\begin{equation}
{}{}\beta _{4}= 
\begin{array}{|lllll|}
\cdot & 1 & \cdot & \cdot & \cdot \\ 
1 & \cdot & \cdot & \cdot & \cdot \\ 
\cdot & \cdot & \cdot & \cdot & \cdot \\ 
\cdot & \cdot & \cdot & \cdot & \cdot \\ 
\cdot & \cdot & \cdot & \cdot & \cdot
\end{array}
\,,\beta _{1}= 
\begin{array}{|lllll|}
\cdot & \cdot & 1 & \cdot & \cdot \\ 
\cdot & \cdot & \cdot & \cdot & \cdot \\ 
1 & \cdot & \cdot & \cdot & \cdot \\ 
\cdot & \cdot & \cdot & \cdot & \cdot \\ 
\cdot & \cdot & \cdot & \cdot & \cdot
\end{array}
\,,\beta _{2}= 
\begin{array}{|lllll|}
\cdot & \cdot & \cdot & 1 & \cdot \\ 
\cdot & \cdot & \cdot & \cdot & \cdot \\ 
\cdot & \cdot & \cdot & \cdot & \cdot \\ 
1 & \cdot & \cdot & \cdot & \cdot \\ 
\cdot & \cdot & \cdot & \cdot & \cdot
\end{array}
\,,\beta _{3}= 
\begin{array}{|lllll|}
\cdot & \cdot & \cdot & \cdot & 1 \\ 
\cdot & \cdot & \cdot & \cdot & \cdot \\ 
\cdot & \cdot & \cdot & \cdot & \cdot \\ 
\cdot & \cdot & \cdot & \cdot & \cdot \\ 
1 & \cdot & \cdot & \cdot & \cdot
\end{array}
\end{equation}
After the integration over $\psi $ and $\bar{\psi}$ in Eq.~(1) we get: 
\begin{eqnarray}
&&Z_{DKP}(\bar{J},J,J_{\mu })=Z_{0}\int_{\beta }DA_{\mu }(x)\exp \left\{
-\int_{\beta }d^{4}x\left[ \frac{1}{4}F_{\mu \nu }^{2}+\frac{1}{2\alpha }%
(\partial _{\mu }A_{\mu })^{2}+J_{\mu }A_{\mu }\right. \right.  \nonumber \\
&&\qquad \qquad \left. \left. {}+\mbox{Tr}\ln S(x,x,A)\right] ^{\beta
}-\int_{\beta }d^{4}xd^{4}y\bar{J}(x)S(x,y,A)J(y)\right\} .
\end{eqnarray}
Here 
\begin{equation}
S(x,y,A)=\left( \beta _{\mu }D_{\mu }+m\right) ^{-1}\delta ^{4}(x-y)
\end{equation}
is the GF of a DKP particle in external field $A_{\mu }(x)$; the term $%
\mbox{Tr}\ln S(x,x,A)$ gives rise to all vacuum perturbations diagrams. This
term can be transformed into the following component form: 
\begin{eqnarray}
&&\det S(x,y,A)=\int_{\beta }D\psi D\bar{\psi}\exp \left\{ -\int_{\beta
}d^{4}x\bar{\psi}\left( \beta _{\mu }D_{\mu }+m\right) \psi \right\} = 
\nonumber \\
&&\kern-10pt{}=\int_{\beta }\prod_{\mu =1}^{4}D\phi _{\mu }D\phi _{\mu
}^{\ast }D\phi D\phi \exp \biggl\{-\int_{\beta }d^{4}x\bigl(\phi ^{\ast
}D_{\mu }\phi _{\mu }+\phi _{\mu }^{\ast }D_{\mu }\phi +m(\phi \phi ^{\ast
}+\phi _{\mu }\phi _{\mu }^{\ast }\bigr)\biggr\}.  \nonumber \\
&&
\end{eqnarray}
Now let us integrate over $\phi _{\mu }$ and $\phi _{\mu }^{\ast }$. We get: 
\begin{eqnarray}
&&\det S(x,y,A)=\det G(x,y,A)=\exp \mbox{Tr}\ln G(x,x,A)=  \nonumber \\
&&\qquad \qquad \frac{1}{m}\int_{\beta }D\phi D\phi ^{\ast }\exp \left\{ -%
\frac{1}{m}\int_{\beta }d^{4}x\phi ^{\ast }\left( -D_{\mu }^{2}+m^{2}\right)
\phi \right\} ,
\end{eqnarray}
where 
\begin{equation}
G(x,y,A)=\left( -D_{\mu }^{2}+m^{2}\right) ^{-1}\delta ^{4}(x-y)
\end{equation}
is the GF of the KGF equation in the case of external field $A_{\mu }(x)$.
Thus, we conclude from Eqs.~(7--9) that all many-photon GF (not only matrix
elements of S-matrix for \textit{real\/} photons) coincide in DKP and KGF
statistical theories\footnote{%
Strictly speaking, the scalar fields $\phi (x)$ in DKP and $\varphi (x)$ in
KGF theory are related by the following equation: 
\[
\varphi (x)=\frac{1}{\sqrt{m}}\phi (x) 
\]
}. This concludes the proof of equivalence for many-photon GF in KGF and DKP
statistical theories.

\section{Polarization Operator in One-Loop Approximation}

Polarization operator in KGF statistical theory for charged spin-zero scalar
particles in one-loop approximation has the form\footnote{%
The last term in Eq.~(10) appears due to the term $e^{2}\int A_{\mu
}^{2}(x)\phi ^{*}(x)\phi (x)d^{4}x$ in the Lagrangian of the KGF theory.} 
\begin{equation}
\Pi _{\mu \nu }^{k}(k)=-\frac{e^{2}}{(2\pi )^{3}\beta }\sum_{p_{4}}\int d%
\mathbf{p}\left( \frac{(2p+k)_{\mu }(2p+k)_{\nu }}{%
(p^{2}+m^{2})((p+k)^{2}+m^{2})}-\frac{2\delta _{\mu \nu }}{p^{2}+m^{2}}%
\right) ,
\end{equation}
where 
\begin{equation}
p^{2}=p_{4}^{2}+\mathbf{p}^{2};\quad p_{4}=\frac{2\pi n}{\beta };\quad
-\infty <n<+\infty .
\end{equation}
The term proportional to $\delta _{\mu \nu }$ in Eq.~(10) is important in
the proof of transversality of $\Pi _{\mu \nu }$ ($k_{\mu }\Pi _{\mu \nu
}(k)=0$). However, this term does not contribute to $\Pi _{\mu \nu }$ after
the renormalization. \noindent In DKP theory, the one-particle GF in
momentum space is: 
\begin{eqnarray}
&&G(\hat{p})=-\frac{1}{m}\left( \frac{i\hat{p}(i\hat{p}+m)}{p^{2}+m^{2}}%
+1\right) ,  \nonumber \\
&&\hat{p}=\beta _{\mu }p_{\mu }.
\end{eqnarray}
It is easy to check that 
\begin{equation}
(i\hat{p}-m)G(\hat{p})=1.
\end{equation}
Using Eqs.~(12)--(13) we obtain the polarization operator in DKP theory (in $%
e^{2}$-ap\-prox\-i\-ma\-ti\-on): 
\begin{eqnarray}
&&\Pi _{\mu \nu }^{D}(k)=\frac{e^{2}}{m^{2}(2\pi )^{3}\beta }\mbox{Tr}%
\sum_{p_{4}}\int d\mathbf{p}\,\beta _{\mu }G(\hat{p}+\hat{k})\beta _{\nu }G(%
\hat{p}) \\
&&\kern-17pt{}=-\frac{e^{2}}{(2\pi )^{3}\beta }\sum_{p_{4}}\int d\mathbf{p}%
\left( \frac{(2p+k)_{\mu }(2p+k)_{\nu }}{(p^{2}+m^{2})((p+k)^{2}+m^{2})}-%
\frac{\delta _{\mu \nu }}{p^{2}+m^{2}}-\frac{\delta _{\mu \nu }}{%
(p+k)^{2}+m^{2}}+\frac{\delta _{\mu \nu }}{m^{2}}\right) .  \nonumber
\end{eqnarray}
The last term $\sim \delta _{\mu \nu }$ in DKP theory breaks the gauge
invariance\footnote{%
See Bernard's work \cite{Bernard}.}, but disappears after the
renormalization. It is easy to show that after the substitution $%
(p+k)\leftrightarrow p$ in the regularization term $\delta _{\mu \nu
}((p+k)^{2}+m^{2})^{-1}$, it will be equal to $\delta _{\mu \nu
}(p^{2}+m^{2})^{-1}$. This coincidence of $\Pi _{\mu \nu }^{K}$ and $\Pi
_{\mu \nu }^{D}$ in one-loop approximation confirms the general proof given
in Section 2, see Eqs.~(8)--(9)\footnote{%
One may note that $\Pi _{\mu \nu }^{D}$ given by Eq.~(8.23) in~\cite{Proc}
does not coincide with our Eq.~(14), breaking the equivalence.}.

The $\Pi _{\mu \nu }(k)$ tensor has the form \cite{Fainberg}, \cite{FPV}: 
\begin{equation}
\Pi _{\mu \nu }=(k_{\mu }k_{\nu }-k^{2}\delta _{\mu \nu })\Pi (k^{2}).
\end{equation}
In quantum statistics, $\Pi _{\mu \nu }$ depends on the two vectors: $k_{\mu
}$ and $u_{\mu }$, this latter is the single vector of medium velocity.
Thus, in the general case (see [3], p.~75) 
\begin{eqnarray}
&&\Pi _{\mu \nu }=\left( \delta _{\mu \nu }-\frac{k_{\mu }k_{\nu }}{k^{2}}%
\right) A_{1}+\left( u_{\mu }u_{\nu }-\frac{k_{\mu }u_{\nu }(ku)}{k^{2}}-%
\frac{k_{\nu }u_{\mu }(ku)}{k^{2}}+\frac{k_{\mu }k_{\nu }(ku)^{2}}{k^{4}}%
\right) A_{2}  \nonumber \\
&&\qquad \qquad {}\equiv \Phi _{\mu \nu }^{1}A_{1}+\Phi _{\mu \nu }^{2}A_{2}.
\label{ec16}
\end{eqnarray}
Introducing the notation (for any approximation) 
\begin{eqnarray}
&&a_{1}\equiv \Pi _{\mu \mu }=3A_{1}+\lambda A_{2} \\
&&a_{2}\equiv u_{\mu }\Pi _{\mu \nu }u_{\nu }=\lambda (A_{1}+\lambda
A_{2}),\qquad \lambda =\left( 1-\frac{(ku)^{2}}{k^{2}}\right) ,
\end{eqnarray}
we get: 
\begin{equation}
A_{1}=\frac{1}{2}\left( a_{1}-\frac{1}{\lambda }a_{2}\right) ,\qquad A_{2}=%
\frac{1}{2\lambda }\left( -a_{1}+\frac{3}{\lambda }a_{2}\right) .
\end{equation}
If the system is at rest\footnote{%
One can show that Eq.~(\ref{ec16}), strictly speaking, is satisfied only if
the system is at \textit{rest\/} ($\mathbf{u}=0,u_{4}=1$), see~\cite{Proc},
Chapter \uppercase\expandafter{\romannumeral 2}, sect.~7.}, 
\begin{equation}
\lambda =\left( 1-\frac{k_{4}^{2}}{k^{2}}\right) ,
\end{equation}
and 
\begin{equation}
a_{2}=\Pi _{44}=\left( 1-\frac{k_{4}^{2}}{k^{2}}\right) A_{1}+\left( 1-\frac{%
k_{4}^{2}}{k^{2}}\right) ^{2}A_{2}.
\end{equation}
It is convenient to represent $a_{1},a_{2}$ in the form: 
\begin{equation}
a_{i}=a_{i}^{R}+a_{i}^{\beta }.
\end{equation}
Here $a_{i}^{R}$ are the parts which do not depend on $\beta $ and which
must be renormalizable; $a_{i}^{\beta }$ depend on $\beta $. It is important
that when the temperature is zero 
\begin{equation}
\lim_{\beta \rightarrow \infty }a_{i}^{\beta }=0.
\end{equation}
Now we can write the Eqs.~(\ref{ec16}) in the following form: 
\begin{equation}
\Pi _{\mu \nu }=\frac{1}{2}\Phi _{\mu \nu }^{1}\left( a_{1}^{R}-{\frac{1}{%
\lambda }}a_{2}^{R}+a_{1}^{\beta }-{\frac{1}{\lambda }}a_{2}^{\beta }\right)
+\frac{1}{2\lambda }\Phi _{\mu \nu }^{2}\left( -a_{1}^{R}+{\frac{3}{\lambda }%
}a_{2}^{R}-a_{1}^{\beta }+{\frac{3}{\lambda }}a_{2}^{\beta }\right) .
\end{equation}
The term $\sim \Phi _{\mu \nu }^{2}$ must vanish in the limit $\beta
\rightarrow \infty $. Therefore in this limit we obtain $\Pi _{\mu \nu }$ of
the Euclidean quantum field theory. So far as $a_{1}^{R}$ and $a_{2}^{R}$ do
not depend on $\beta $, we conclude that (after the renormalization) 
\begin{equation}
a_{2}^{R}={\frac{\lambda }{3}}a_{1}^{R}.
\end{equation}
Thus 
\begin{equation}
\lim_{\beta \rightarrow \infty }\Pi _{\mu \nu }={\frac{1}{3}}\Phi _{\mu \nu
}^{1}a_{1}^{R}\qquad \mbox{or}\qquad \Pi _{\mu \mu }=a_{1}^{R}.
\end{equation}
We calculate $a_{1}$ and $a_{2}$ using the general formula for summation
over $p_{4}$ in Eq.~(14). We ignore the terms $\sim \delta _{\mu \nu }$ in
Eqs.~(10) and~(14) since these terms disappear after regularization and
renormalization. 
\begin{eqnarray}
&&a_{1}=\Pi _{\mu \mu }=-\frac{e^{2}}{(2\pi )^{3}\beta }\int d\mathbf{p}%
\sum_{p_{4}}\frac{(2p+k)^{2}}{(p^{2}+m^{2})((p+k)^{2}+m^{2})} \\
&&a_{2}=\Pi _{44}=-\frac{e^{2}}{(2\pi )^{3}\beta }\int d\mathbf{p}%
\sum_{p_{4}}\frac{(2p_{4}+k_{4})^{2}}{(p^{2}+m^{2})((p+k)^{2}+m^{2})}\,,
\end{eqnarray}
where $p_{4}={2\pi n/\beta }$.

The general formula for summation over $\beta $ is\footnote{%
See \cite{Proc}, page 123, supplement 3 and \cite{Rothe}, page 299.} 
\begin{equation}
\kern-20pt{\frac{1}{\beta }}\sum_{n}f({\frac{2\pi n}{\beta }},K)={\frac{1}{%
2\pi }}\int_{-\infty }^{\infty }d\omega f(\omega ,K)+{\frac{1}{2\pi }}%
\int_{-\infty +i\epsilon }^{\infty +i\epsilon }d\omega \frac{f(\omega
,K)+f(-\omega ,K)}{e^{-i\beta \omega }-1}\,.
\end{equation}
The Eq.~(27) can be rewritten in the following form: 
\begin{eqnarray}
&&a_{1}=-\frac{e^{2}}{(2\pi )^{3}\beta }\int d\mathbf{p}\sum_{p_{4}}\frac{%
4p^{2}+4pk+2k^{2}+4m^{2}-(4m^{2}+k^{2})}{(p^{2}+m^{2})((p+k)^{2}+m^{2})} 
\nonumber \\
&&{}=-\frac{e^{2}}{(2\pi )^{3}\beta }\int d\mathbf{p}\sum_{p_{4}}\left\{ -%
\frac{4m^{2}+k^{2}}{(p^{2}+m^{2})((p+k)^{2}+m^{2})}+\frac{2}{p^{2}+m^{2}}+%
\frac{2}{(p+k)^{2}+m^{2}}\right\}  \nonumber \\
&&\qquad \qquad {}\Rightarrow +\frac{e^{2}}{(2\pi )^{3}\beta }\int d\mathbf{p%
}\sum_{p_{4}}\frac{4m^{2}+k^{2}}{(p^{2}+m^{2})((p+k)^{2}+m^{2})},
\end{eqnarray}
where we omit the last two terms which vanish after renormalization; $%
p^{2}=p_{4}^{2}+\mathbf{p}^{2}$.

Introducing the Feynman's parameter $x$ into Eq.~(30), we obtain: 
\begin{equation}
a_{1}(k^{2})=-\frac{e^{2}}{(2\pi )^{3}\beta }(4m^{2}+k^{2})\int d\mathbf{p}%
\frac{\partial }{\partial m^{2}}\int_{0}^{1}dx\sum_{p_{4}}\frac{1}{\left[
(p^{2}+m^{2})+\frac{k^{2}}{4}(1-x^{2})\right] }.
\end{equation}

To get the contribution to $a_{1}(k^{2})$ which does not depend on $\beta $
we must use only the first term of Eq.~(29): 
\begin{equation}
a_{1}^{R}(k^{2})=-\frac{e^{2}}{(2\pi )^{3}}(4m^{2}+k^{2})\int d\mathbf{p}%
\frac{\partial }{\partial m^{2}}\int_{0}^{1}dx\frac{1}{2\pi }\int_{-\infty
}^{\infty }\frac{d\omega }{\left[ \omega ^{2}+\mathbf{p}^{2}+m^{2}+\frac{%
k^{2}}{4}(1-x^{2})\right] }.
\end{equation}

Closing the integration contour at infinity in the upper half-plane, we
find: 
\begin{equation}
a_{1}^{R}(k^{2})=-\frac{e^{2}}{(2\pi )^{3}}(4m^{2}+k^{2})\int^{l}d\mathbf{p}%
\frac{\partial }{\partial m^{2}}\int_{0}^{1}dx\frac{1}{2\left[ \mathbf{p}%
^{2}+m^{2}+\frac{k^{2}}{4}(1-x^{2})\right] ^{1/2}},
\end{equation}
where $l$ is the momentum cut-off.

After the integration over $x$ and renormalization 
\begin{equation}
a_{1}^{R}(k^{2})\rightarrow a_{1}^{R}(k^{2})-a_{1}^{R}(0)-k^{2}{\frac{%
\partial }{\partial k^{2}}}a_{1}^{R}(0),
\end{equation}
we obtain: 
\begin{equation}
a_{1}^{R}(k^{2})=-\frac{e^{2}k^{4}}{16\pi ^{2}}\int_{4m^{2}}^{\infty }\frac{%
dz^{2}\left( 1-\frac{4m^{2}}{z^{2}}\right) ^{3/2}}{z^{2}(z^{2}+k^{2})}.
\end{equation}
In the limit $\beta \rightarrow \infty $ we get the Euclidean expression for 
$\Pi _{\mu \nu }$ (see Eq.~(26)): 
\begin{equation}
\lim_{\beta \rightarrow \infty }\Pi _{\mu \nu }=\left( -\frac{k_{\mu }k_{\nu
}}{k^{2}}+\delta _{\mu \nu }\right) \left( \frac{e^{2}}{48\pi ^{2}}\right)
k^{4}\int_{4m^{2}}^{\infty }\frac{dz\left( 1-\frac{4m^{2}}{z^{2}}\right)
^{3/2}}{z^{2}(z^{2}+k^{2})}.
\end{equation}
This expression for $\Pi _{\mu \nu }$ also follows from Eq.~(11) in \cite
{FPV}, where the photon GF has been calculated in DKP theory using
dispersion approach.

One can easily find $a_{2}^{R}$ from the Eqs.~(25),(35): 
\begin{equation}
a_{2}^{R}(k^{2})=\frac{\lambda }{3}a_{1}^{R}=+\frac{e^{2}k^{2}}{48\pi ^{2}}%
(k_{4}^{2}-k^{2})\int_{4m^{2}}^{\infty }\frac{dz\left( 1-\frac{4m^{2}}{z^{2}}%
\right) ^{3/2}}{z^{2}(z^{2}+k^{2})}.
\end{equation}
One can write the expression\footnote{%
Some details are given in the Appendix.} for $a_{1}^{\beta }$ and $%
a_{2}^{\beta }$ ($\mu \neq 0$): 
\begin{eqnarray}
&&a_{1}^{\beta }=\frac{e^{2}}{16\pi ^{2}}(4m^{2}+k^{2})\int_{0}^{\infty }%
\frac{p\,dp}{E|\mathbf{k}|}\left( e^{\beta (E-\mu )}-1\right) ^{-1}\ln \frac{%
(k^{2}+2p\mathbf{k})^{2}+4E^{2}k_{4}^{2}}{(k^{2}-2p\mathbf{k}%
)^{2}+4E^{2}k_{4}^{2}}  \nonumber \\
&& \\
&&a_{2}^{\beta }=\frac{e^{2}}{16\pi ^{2}}\int \frac{p^{2}\,dp}{Ep|\mathbf{k}|%
}\left( e^{\beta (E-\mu )}-1\right) ^{-1}\biggl\{(E^{2}-k_{4}^{2})\ln \frac{%
(k^{2}+2p\mathbf{k})^{2}+4E^{2}k_{4}^{2}}{(k^{2}-2p\mathbf{k}%
)^{2}+4E^{2}k_{4}^{2}}  \nonumber \\
&&{}+2iEk_{4}\ln \frac{(k^{2}+2iEk_{4})^{2}-4p^{2}\mathbf{k}^{2}}{%
(k^{2}-2iEk_{4})^{2}-4p^{2}\mathbf{k}^{2}}\biggr\},
\end{eqnarray}
where 
$$
E=(p^{2}+m^{2})^{1/2}.\eqno{(39a)} 
$$

Some details of calculations can be found in the Appendix.

\section{Conclusions}

We have proved the equivalence of photon GF in DKP and KGF statistical
theories (Section \textbf{2}), and carried out calculations of polarization
operator in one-loop approximation to illustrate the equivalence (Section 
\textbf{3}). It would be interesting to generalize the proof of equivalence
for GF of many non-Abelian gluons in statistical quantum field DKP and KGF
theories (see \cite{FP1}).

The generalization of the proof of equivalence for photons GF in DKP and KGF
statistical theories to the case of charged vector fields can also be made,
however this proof will have a formal character due to
non-renormaliza-bility of the theory.

\section*{Acknowledgments}

One of us (V.Ya.F) thanks Prof. I.V.Tyutin and Prof. A.E.Shabad for useful
discussions. B.M.P. and J.S. Valverde are grateful for R. Casana's comments.
This work was supported by FAPESP (V.Ya.F., grant 01/12585-6; B.M.P., grant
02/00222-9; J.S.V. full support grant 00/03812-6), RFBR/Russia (V.Ya.F.,
grant 02-02-16946 and 02-01-00556), LSS-1578.2003.2 and CNPq/Brazil (B.M.P.).

\section*{Appendix}

1. Let us consider the derivation of Eq.~(35) in more detail. We start from
Eq.~(33), which can be rewritten in the form:

\begin{eqnarray*}
&&a_{1}^{R}(k^{2})=-\frac{e^{2}}{(2\pi )^{3}}4\pi (m^{2}+k^{2})\frac{%
\partial }{\partial m^{2}}\int_{0}^{1}dx\int_{0}^{l}p^{2}\,dp\left(
p^{2}+m^{2}+\frac{k^{2}}{4}(1-x^{2})\right) ^{-1/2} \\
&&\kern-18pt{}=-\frac{e^{2}}{(2\pi )^{2}}\sqrt{\frac{4}{k^{2}}}%
\int_{0}^{l}p^{2}\,dp(m^{2}+k^{2})\frac{\partial }{\partial m^{2}}%
\int_{0}^{1}dx(\alpha +1-x^{2})^{-1/2};\mbox{ where }\alpha =\frac{%
4(p^{2}+m^{2})}{k^{2}}, \\
&&\qquad {}=-\frac{e^{2}}{(2\pi )^{2}}\sqrt{\frac{4}{k^{2}}}%
\int_{0}^{l}p^{2}\,dp(4m^{2}+k^{2})\left( \frac{4}{k^{2}}\right) {\frac{%
\partial }{\partial \alpha }}\int_{0}^{\frac{1}{\sqrt{1+\alpha }}}\frac{dy}{%
(1-y^{2})^{1/2}} \\
&&\quad {}=-\left( \frac{e^{2}}{8\pi ^{2}}\right) \int_{0}^{l}\frac{p^{2}\,dp%
}{E}\frac{4m^{2}+k^{2}}{k^{2}+4m^{2}+p^{2}}\Rightarrow 
\mbox{\rm after
renormalization, see Eq.(34)} \\
&&\qquad \qquad {}=-\frac{e^{2}}{16\pi ^{2}}(k^{2})^{2}\int_{4m^{2}}^{\infty
}\frac{dz^{2}\left( 1-\frac{4m^{2}}{z^{2}}\right) ^{3/2}}{z^{2}(z^{2}+k^{2})}%
.
\end{eqnarray*}

e-Proc. http://www.sbf1.if.usp.br/eventos/enfpc/xx/procs/res127/

\end{document}